\documentstyle[12pt]{article}

\newcommand{\F}{\noindent}
\newcommand{\qqq}{\qquad\qquad}

\newcommand{\q}{\quad}
\newcommand{\SP}{\smallskip}
\newcommand{\MP}{\medskip}
\newcommand{\BP}{\bigskip}

\newcommand{\beq}{\begin{eqnarray}}
\newcommand{\ene}{\end{eqnarray}}
\newcommand{\beqn}{\begin{eqnarray*}}
\newcommand{\enen}{\end{eqnarray*}}

\newcommand{\R}{{\mbox{\bf R}}}
\newcommand{\N}{{\mbox{\bf N}}}
\newcommand{\HH}{{\cal H}}
\newcommand{\UU}{{\cal U}}
\newcommand{\OO}{{\cal O}}
\newcommand{\la}{{\lambda}}

\newcommand{\Ltnn}{{L^2(R^{3n})}}

\newcommand{\tT}{{\widetilde T}}
\newcommand{\tH}{{\widetilde H}}

\newcommand{\SSS}{{\cal{S}}}

\newcommand{\eq}[1]{(\ref{#1})}

\newenvironment{namelist}[1]{%
\begin{list}{}
{
\settowidth{\labelwidth}{#1}
\setlength{\leftmargin}{1.1\labelwidth}}
}{%
\end{list}}

\Large

\begin{document}
\begin{flushright}
KIMS-2001-10-12\\
gr-qc/0110066
\end{flushright}

\BP

\begin{center}
\Large

{\bf Local Time and the Unification of Physics}

{\bf Part II. Local System}

\normalsize

\vskip12pt

Hitoshi Kitada

Department of Mathematical Sciences

University of Tokyo

Komaba, Meguro, Tokyo 153, Japan

E-mail: kitada@ms.u-tokyo.ac.jp
\vskip6pt

%
%
%
%
%
%
\vskip10pt

October 12, 2001
\vskip2pt

\end{center}

\vskip24pt

\normalsize

\leftskip24pt
\rightskip24pt

\small

\noindent
{\it Abstract}:
As a continuation of Part I \cite{[Ki-Fl]}, a more precise formulation of local time and local system is given. The observation process is reflected in order to give a relation between the classical physics for centers of mass of local systems and the quantum mechanics inside a local system. The relation will give a unification of quantum mechanics and general relativity in some cases. The existence of local time and local motion is proved so that the stationary nature of the universe is shown to be consistent with the local motion.

\leftskip0pt
\rightskip0pt

\vskip 24pt

\large
\noindent
{\bf V. Definition of Local System}

\vskip12pt

\normalsize

\noindent
As announced in Part I \cite{[Ki-Fl]}, we begin in this part II with stating a precise definition of local clock and local time, where as we have discussed in Part I, local clock is the local system itself in that every existence is clocking. (See section IV \cite{[Ki-Fl]}, especially recall the statement: In this sense, ``clocking" is the natural activity of any local system.) Thus the purpose of this section is the definition of local system.

To do so, we begin
 with introducing a stationary universe $\phi$. By nature what is called the universe must be a closed universe, within which
 is all. We characterize it by a certain
 quantum-mechanical condition.

Let $\HH$ be a separable Hilbert space, and set
$$
\UU=\{\phi\}=\bigoplus_{n=0}^\infty \left(\bigoplus_{\ell=0}^\infty
\HH^n  \right) \q (\HH^n=\underbrace{\HH\otimes\cdots\otimes
\HH}_{\scriptsize \mbox{$n$ factors}}).
$$
$\UU$ is called a Hilbert space of possible universes. An element 
$\phi$ of $\UU$ is called a universe and is of the form of an
 infinite matrix $(\phi_{n\ell})$ with components $\phi_{n\ell} 
\in \HH^n$. $\phi=0$ means $\phi_{n\ell}=0$ for all $n, \ell$.

Let $\OO=\{ S\}$ be the totality of the selfadjoint operators $S$
 in $\UU$ of the form $S\phi=(S_{n\ell}\phi_{n\ell})$ for
$\phi=(\phi_{n\ell})\in{\cal D}(S)\subset\UU$, where each component
 $S_{n\ell}$ is a selfadjoint operator in $\HH^n$. Our characterization of the universe $\phi$ is the
 following condition.

{\bf Axiom 1.}\ There is a selfadjoint operator $H\in\OO$ in 
$\UU$ such that for some $\phi\in \UU-\{0\}$ and $\la\in \R$
\beq
H\phi\approx\la \phi \label{eq1}
\ene
in the following sense: Let $F_n$ be a finite subset of 
${\N}=\{1,2,\cdots\}$ with $\sharp(F_n)(=$ the number of elements in
 $F_n)=n$ and let $\{ F_n^\ell\}_{\ell=0}^\infty$ be the totality of
 such $F_n$ (note: the set $\{ F_n^\ell\}_{\ell=0}^\infty$ is countable). Then the formula \eq{eq1} in the above means that there are
 integral sequences $\{n_k\}_{k=1}^\infty$ and 
$\{ \ell_k\}_{k=1}^\infty$ and a real sequence 
$\{\la_{n_k \ell_k}\}_{k=1}^\infty$
such that 
$F_{n_k}^{\ell_k}\subset F_{n_{k+1}}^{\ell_{k+1}}$; 
$\bigcup_{k=1}^\infty F_{n_k}^{\ell_k} = \N$;
\beq
H_{n_k\ell_k}\phi_{n_k\ell_k}=\la_{n_k\ell_k}\phi_{n_k\ell_k},
\q \phi_{n_k\ell_k}\ne0,\q k=1,2,3,\cdots; \label{eq2}
\ene
and
$$
\la_{n_k\ell_k}\to\la\q {\mbox{as}}\q k\to\infty.
$$

 $H$ is an infinite matrix $(H_{n\ell})$ of selfadjoint operators
 $H_{n\ell}$ in $\HH^n$. Axiom 1 asserts that this matrix converges
 in the sense of \eq{eq1} on our universe $\phi$. We remark that our
 universe $\phi$ is not determined uniquely by this condition.

 The universe as a state $\phi$ is a whole, within which is all.
 As such a whole, the state $\phi$ can follow the two ways: The one
 is that $\phi$ develops along a global time $T$ in the grand
 universe $\UU$ under a propagation $\exp(-iTH)$, and another is
 that $\phi$ is a bound state of $H$. If there were such a global
 time $T$ as in the first case, all phenomena had to develop along
 that global time $T$, and the locality of time would be lost. We
 could then {\it not} construct a notion of local times compatible
 with general theory of relativity. The only one possibility is
 therefore to adopt the stationary universe $\phi$ of Axiom 1.

\BP

The following axiom asserts the existence of configuration and
 momentum operators and that the canonical commutation relation
 between them holds. This is a basis of our definition of time,
 where configuration and momentum are given first, and then local
 times are defined in each local system of finite number of 
quantum-mechanical particles.

\MP

{\bf Axiom 2.}\ Let $n\ge 1$ and $F_{n+1}$ be a finite subset of 
${\N}=\{1,2,\cdots\}$ with $\sharp(F_{n+1})=n+1$. Then for any 
$j\in F_{n+1}$, there are selfadjoint operators 
$X_j =(X_{j 1},X_{j2},X_{j3})$ and $P_j =(P_{j1},P_{j2},P_{j3})$ in
$\HH^n$, and constants $m_j>0$ such that
$$
[X_{j\ell},X_{k m}]=0,\q [P_{j\ell},P_{k m}]=0,
\q [X_{j\ell},P_{k m}]=i\delta_{jk}\delta_{\ell m},
$$
$$
\sum_{j\in F_{n+1}} m_j X_j=0,\q \sum_{j\in F_{n+1}} P_j=0.
$$

The Stone-von Neumann theorem and Axiom 2 specify the space
 dimension (see  \cite{[A-M]}, p.452) as 3 dimension. We identify $\HH^n$ 
with $\Ltnn$ in the following.

\BP

What we intend to mean by the $(n,\ell)$-th component $H_{n\ell}$ 
$(n,\ell\ge 0)$ of $H=(H_{n\ell})$ in Axiom 1 is the usual $N=n+1$
 body Hamiltonian with center of mass removed to accord to the
 requirement $\sum_{j\in F_{n+1}} m_j X_j=0$ in Axiom 2. For the
 local Hamiltonian $H_{n\ell}$ we thus make the following postulate.

\MP

{\bf Axiom 3.}\ Let $n\ge0$ and $F_N$ $(N=n+1)$ be a finite subset
 of ${\N}=\{1,2,\cdots\}$ with $\sharp(F_N)=N$. Let 
$\{ F_N^\ell\}_{\ell=0}^\infty$ be the countable totality of such $F_N$. Then
 the component Hamiltonian $H_{n\ell}$ $(\ell\ge0)$ of $H$ in Axiom
 1 is of the form
$$
H_{n\ell}=H_{n\ell0}+V_{n\ell},\q V_{n\ell}=\sum_{{\scriptstyle \alpha=(i,j)}
\atop{\scriptstyle 1\le i<j<\infty,\ i,j\in F_N^\ell}} V_\alpha(x_\alpha)
$$
on $C_0^\infty(\R^{3n})$, where $x_\alpha=x_i-x_j$ $(\alpha=(i,j))$ with $x_i$ being
 the position vector of the $i$-th particle, and 
$V_\alpha(x_\alpha)$ is a real-valued measurable function of 
$x_\alpha\in \R^3$ which is $H_{n\ell0}$-bounded with 
$H_{n\ell0}$-bound of $V_{n\ell}$ less than 1.
$H_{n\ell0}=H_{(N-1)\ell0}$ is the free Hamiltonian of the 
$N$-particle system, which has the form like
$$
\sum_{\ell=1}^n\sum_{k=1}^3 \frac{1}{2\mu_\ell}
\frac{\partial^2}{\partial x_{\ell k}^2}\quad \mbox{with } \mu_\ell>0 \mbox{ being reduced mass}.
$$

This axiom implies that $H_{n\ell}=H_{(N-1)\ell}$ is uniquely
 extended to a selfadjoint operator bounded from below in 
$\HH^n=\HH^{N-1} =L^2(\R^{3(N-1)})$ by the Kato-Rellich theorem.

We do not include vector potentials in the Hamiltonian 
$H_{n\ell}$ of Axiom 3, for we take the position that what 
is elementary is the electronic charge, and the magnetic 
forces are the consequence of the motions of charges.

\BP

Let $P_H$ denote the orthogonal projection onto the space of
bound states for a selfadjoint operator $H$. We call the set of all states
 orthogonal to the space of bound states a scattering space, and its element as a scattering state. Let $\phi=(\phi_{n\ell})$ with 
$\phi_{n\ell}=\phi_{n\ell} (x_1,\cdots,x_n)\in \Ltnn$ 
be the universe in Axiom 1, and let $\{n_k\}$ and $\{\ell_k\}$ be
 the sequences specified there. Let $x^{(n,\ell)}$ denote the
 relative coordinates of $n+1$ particles in $F_{n+1}^\ell$.
\MP


\F
{\bf Definition 1.}
\begin{namelist}{888}
\item[(1)]
We define $\HH_{n\ell}$ as the sub-Hilbert space of $\HH^n$
 generated by the functions $\phi_{n_k\ell_k}$ $(x^{(n,\ell)},y)$ of
 $x^{(n,\ell)}\in \R^{3n}$ with regarding $y\in\R^{3(n_k-n)}$ as a
 parameter, where $k$ moves over a set $\{k\ |\ n_k\ge n, 
F_{n+1}^\ell\subset F_{n_k+1}^{\ell_k}, k\in \N\}$.
\item[(2)]
$\HH_{n\ell}$ is called a {\it local universe} of  $\phi$.
\item[(3)]
$\HH_{n\ell}$ is said to be non-trivial if 
$(I-P_{H_{n\ell}})\HH_{n\ell}\ne\{0\}$.
\end{namelist}

The total universe $\phi$ is a single element in  $\UU$. The local
 universe $\HH_{n\ell}$ can be richer and may have elements more
 than one. This is because we consider the subsystems of the
universe consisting of a finite number of particles. These
 subsystems receive the influence from the other particles of
 infinite number outside the subsystems, and may vary to constitute
 a non-trivial subspace $\HH_{n\ell}$. We will return to this point in section VII.

\hyphenation{Coulomb}

We can now define local system.

\BP

\F
{\bf Definition 2.}
\begin{namelist}{888}
\item[(1)]
The restriction of $H$ to $\HH_{n\ell}$ is also denoted by the same
 notation  $H_{n\ell}$ as the $(n,\ell)$-th component of $H$.
\item[(2)]
We call the pair
$(H_{n\ell},\HH_{n\ell})$  a local system.
\item[(3)]
The unitary group
$e^{-itH_{n\ell}}$ $(t\in \R^1)$ on $\HH_{n\ell}$ is called the 
{\it proper clock} of the local system $(H_{n\ell},\HH_{n\ell})$, if
 $\HH_{n\ell}$ is non-trivial: 
$(I-P_{H_{n\ell}})\HH_{n\ell}\ne \{0\}$.
(Note that the clock is defined only for $N=n+1\ge 2$, since 
$H_{0\ell}=0$ and $P_{H_{0\ell}}=I$.)
\item[(4)]
The universe $\phi$ is called {\it rich} if $\HH_{n\ell}$ equals
$\HH^n=L^2(\R^{3n})$ for all $n\ge 1$, $\ell\ge0$. For a rich
 universe $\phi$, $H_{n\ell}$ equals the $(n,\ell)$-th component of
 $H$.
\end{namelist}


\F
{\bf Definition 3.}
\begin{namelist}{888}
\item[(1)]
The parameter $t$ in the exponent of the proper clock 
$e^{-itH_{n\ell}}=e^{-itH_{(N-1)\ell}}$ of a local system 
$(H_{n\ell},\HH_{n\ell})$ is called the (quantum-mechanical) 
{\it proper time} or {\it local time} of the local system 
$(H_{n\ell}, \HH_{n\ell})$, if $(I-P_{H_{n\ell}})\HH_{n\ell}\ne
 \{0\}$. 
\item[(2)]
This time $t$ is denoted by $t_{(H_{n\ell},\HH_{n\ell})}$ indicating
 the local system under consideration.
\end{namelist}

This definition is a one reverse to the usual definition of the
 motion or dynamics of the $N$-body quantum systems, where the time
 $t$ is given {\it a priori} and then the motion of the particles is
 defined by $e^{-itH_{(N-1)\ell}}f$ for a given initial state $f$ of
 the system.

{\it Time} is thus defined only for local systems
$(H_{n\ell},\HH_{n\ell})$  and is determined by the associated
 proper clock $e^{-itH_{n\ell}}$. Therefore there are infinitely
 many number of times $t=t_{(H_{n\ell},\HH_{n\ell})}$ each of which
 is proper to the local system $(H_{n\ell},\HH_{n\ell})$. In this
 sense time is a local notion. There is no time for the total
 universe $\phi$ in Axiom 1, which is a bound state of the total
Hamiltonian $H$ in the sense of the condition \eq{eq1} of 
Axiom 1.

\BP

 To see the meaning of our definition of time, we quote a theorem
 from \cite{[En]}. To state the theorem we introduce some notations
 concerning the local system 
$(H_{n\ell},\HH_{n\ell})$, assuming that the universe $\phi$ is 
rich: Let $b=(C_1,\cdots,C_{\sharp(b)})$ be a decomposition
 of the set $\{1,2,\cdots,N\}$ $(N=n+1)$ into $\sharp(b)$ number of
 disjoint subsets $C_1,\cdots,C_{\sharp(b)}$ of $\{1,2,\cdots,N\}$.
 $b$ is called a cluster decomposition. 
 $H_b=H_{n\ell,b}=H_{n\ell}-I_b=H^b_{n\ell}+T_{n\ell,b}=H^b+T_b$ is
 the truncated Hamiltonian for the cluster decomposition $b$ with 
 $1\le\sharp(b)\le N$, where $I_b$ is the sum of the intercluster
 interactions between various two different clusters in $b$, and 
 $T_b$ is the sum of the intercluster free energies
 among various clusters in $b$.  $x_b\in \R^{3(\sharp(b)-1)}$ 
is the intercluster coordinates among the centers of mass of the
 clusters in $b$, while $x^b\in \R^{3(N-\sharp(b))}$ denotes the
 intracluster coordinates inside the clusters of $b$ so that 
$x\in \R^{3n}=\R^{3(N-1)}$ is expressed as $x=(x_b,x^b)$.
 Note that $x^b$ is decomposed as $x^b=(x^b_1,\cdots,x^b_{\sharp(b)})$,
 where each $x^b_j\in \R^{3(\sharp(C_j)-1)}$ is the internal
 coordinate of the cluster $C_j$, describing the configuration of
 the particles inside $C_j$. The operator $H^b$ is accordingly
 decomposed as $H^b=H_1+\cdots+H_{\sharp(b)}$, and each component
 $H_j$ is defined in the space 
$\HH^b_j=L^2(\R^{3(\sharp(C_j)-1)}_{x^b_j})$, whose tensor
 product $\HH^b_1\otimes\cdots\otimes\HH^b_{\sharp(b)}$ is the
 internal state space $\HH^b=L^2(\R^{3(N-\sharp(b))}_{x^b})$.
 The free energy $T_b$ is defined in the external space 
$\HH_b=L^2(\R^{3(\sharp(b)-1)}_{x_b})$, and the truncated
 Hamiltonian $H_b=H^b+T_b=I\otimes H^b+T_b\otimes I$ is defined
 in the total space 
$\HH_{n\ell}=\HH_b\otimes \HH^b=L^2(\R^{3(N-1)}_x)$.
$v_b$ is the velocity operator conjugate to the
 intercluster coordinates $x_b$.
 $P_b=P_{H^b}$ is the eigenprojection associated with the subsystem
 $H^b$ of $H$, i.e. the orthogonal projection onto the eigenspace of
 $H^b$, which is defined in $\HH^b$ and extended as $I\otimes P_b$
 to the total space
 $\HH_{n\ell}$. $P_b^M$ is the $M$-dimensional
 partial projection of this eigenprojection $P_b$.
 We define for
 a $k$-dimensional multi-index $M=(M_1,\cdots,M_k)$, $M_j \ge 1$ and
 $k=1,\cdots,N-1$,
$$
{\widehat P} ^M_{k}= \left(I-\sum_{\sharp(b) = k}P_b^{M_k}\right)
\cdots \left(I-\sum_{\sharp(d) = 2} P_d^{M_{2}}\right)
(I-P^{M_1}),
$$
where note that $P^{M_1}=P_a^{M_1}=P_H^{M_1}$ for $\sharp(a)=1$ is
 uniquely determined. We also define for a $\sharp(b)$-dimensional
multi-index $M_{b} = (M_1, \cdots ,$ $ M_{\sharp(b) -1},
M_{\sharp(b)}) = ({\widehat M}_{b}, M_{\sharp(b)})$
$$
{\widetilde P}_{b}^{M_{b}}=P_b^{M_{\sharp(b)}}{\widehat P}_{\sharp(b)
-1}^{{\widehat M}_{b}}, \q 2\le\sharp(b)\le N.
$$
It is clear that
$$
\sum_{2\le\sharp(b)\le N} {\widetilde P}^{M_{b}}_{b} = I - P^{M_1},
$$
provided that the component $M_k$ of $M_{b}$ depends only on the
 number $k$ but  not on  $b$.  In the following we use such 
$M_{b}$'s only. Under these circumstances, the following is known to
 hold.
\BP

\newtheorem{theorem}{Theorem}

\begin{theorem}[\cite{[En]}]
 Let $N=n+1\ge 2$ and let
 $H_{N-1}=H_{n\ell}$ be the Hamiltonian for a local system
 $(H_{n\ell},\HH_{n\ell})$. Let suitable conditions on the
 decay rate for the pair potentials $V_{ij}(x_{ij})$ be 
satisfied {\rm (}see, e.g., Assumption 1 in 
{\rm{\cite{[Ki($N$)]}}}{\rm )}. Let 
$\Vert |x^a|^2 P^M_a\Vert<\infty$ be satisfied for any integer 
$M\ge1$ and cluster decomposition $a$ with $2\le\sharp(a)\le N-1$.
 Let $f \in \HH^{N-1}$.  Then there  is a sequence 
$t_m \to \pm\infty \ ({\mbox{as }} m \to \pm\infty)$ and a sequence 
 $M^m_{b}$ of multi-indices whose components all tend to $\infty$ as
 $m \to \pm \infty$ such that for all cluster decompositions $b$, 
$2\le\sharp(b)\le N$, and 
$ \varphi \in C_0^\infty (\R^{3(\sharp(b) -1)}_{x_b})$
\beq
\Vert \{ \varphi (x_b/t_m) - \varphi (v_b) \} 
{\widetilde P}_{b}^{M^m_{b}}e^{-it_m H_{N-1}} f \Vert \to 0
\label{eq3}
\ene
as $m \to \pm\infty$.
\end{theorem}
\BP

The asymptotic relation \eq{eq3} roughly means that, if we restrict our
 attention to the part ${\widetilde P}_{b}^{M^m_{b}}$ of the evolution 
$e^{-it H_{N-1}}f$, in which the particles inside any cluster of $b$
 are bounded while any two different clusters of $b$ are scattered,
 then the magnitude of quantum-mechanical velocity $v_b=m_b^{-1}p_b$, where $m_b$
 is some diagonal mass matrix, is approximated by the square root of a classical value
 $$
|v_b^{(c)}|^2=\lim_{m\to\pm\infty}
(|v_b|^2{\widetilde P}_{b}^{M^m_{b}}e^{-it_m H_{N-1}}f,
{\widetilde P}_{b}^{M^m_{b}}e^{-it_m H_{N-1}}f)
$$
 asymptotically as $m\to\pm\infty$ and the local time $t$
 of the $N$ body system $H_{N-1}=H_{n\ell}$
 is asymptotically equal to the quotient of the configuration by the
 velocity of the scattered particles (or clusters, exactly 
speaking):
\beq
\frac{\left|x_b\right|}{\left|v_b^{(c)}\right|}\label{eq4}
\ene
on the evolving state ${\widetilde P}_{b}^{M^m_{b}}e^{-it_m H_{N-1}}f$.
This means by $v_b=m_b^{-1}p_b$ that the local time $t$ is
 asymptotically and approximately
 measured if the values of the configurations and momenta for the
 scattered particles of the local system 
$(H_{N-1}, \HH_{N-1})=(H_{n\ell},\HH_{n\ell})$ are given.

\hyphenation{apparent}

We note that the time measured by \eq{eq4} is independent of the choice
 of cluster decomposition $b$ according to Theorem 1. This means that
 $t$ can be taken as a
 common parameter of motion inside the local system, and can be
 called {\it time} of the local system in accordance with the notion of
 `common time' in Newton's sense: ``relative, apparent, and common time,
 is some sensible and external (whether accurate or unequable)
 measure of duration by the means of motion, $\cdots$" (\cite{[New]}, p.6).
 Once we take 
$t$ as our notion of time for the system $(H_{n\ell},\HH_{n\ell})$,
 $t$ recovers the usual meaning of time, by the identity for
 $e^{-itH_{n\ell}}f$ known
 as the Schr\"odinger equation:
$$
\left(\frac{1}{i}\frac{d}{dt}+H_{n\ell}\right)e^{-itH_{n\ell}}f=0.
$$

\hyphenation{spaces}

Time $t=t_{(H_{n\ell},\HH_{n\ell})}$ is a notion defined only in
 relation with  the local system $(H_{n\ell},\HH_{n\ell})$. To other
 local system $(H_{mk},\HH_{mk})$, there is associated other local
 time $t_{(H_{mk},\HH_{mk})}$, and between 
$t=t_{(H_{n\ell},\HH_{n\ell})}$ and $t_{(H_{mk},\HH_{mk})}$,
 there is no relation, and they are completely independent notions.
 In other words, $\HH_{n\ell}$ and $\HH_{mk}$ are different spaces
 unless $n=m$ and $\ell=k$. And even when the two local systems 
$(H_{n\ell},\HH_{n\ell})$ and $(H_{mk},\HH_{mk})$ have a 
non-vanishing common part: $F_{n+1}^\ell\cap F_{m+1}^k\ne\emptyset$,
 the common part constitutes its own local system 
$(H_{pj},\HH_{pj})$, and its local time cannot be compared with
 those of the two bigger systems $(H_{n\ell},\HH_{n\ell})$ and 
$(H_{mk},\HH_{mk})$, because these three systems have different base
 spaces, Hamiltonians, and clocks. More concretely speaking, the
 times are measured through the quotients \eq{eq4} for each system. But
 the $L^2$-representations of the base Hilbert spaces 
$\HH_{n\ell},\HH_{mk},\HH_{pj}$ for those systems are different
 unless they are identical with each other, and the quotient \eq{eq4} has
 incommensurable meaning among these representations.

In this sense, local systems are independent mutually. Also they
 cannot be decomposed into pieces in the sense that the decomposed
 pieces constitute different local systems.

\BP

\newpage


\vskip 24pt

\large
\noindent
{\bf VI. Relativity as Observation}

\vskip12pt

\normalsize

\F
We now see how we can combine relativity and quantum mechanics in our formulation.
\BP

\noindent
{\bf VI.1. Relativity}

\BP

\F
We note that the center of mass of a local system 
$(H_{n\ell},\HH_{n\ell})$ is always at the origin of the space
 coordinate system $x_{(H_{n\ell},\HH_{n\ell})}\in\R^3$ for the
 local system by the requirement: $\sum_{j\in F_{n+1}} m_j X_j=0$ in
 Axiom 2, and that the space coordinate system describes
 just the relative motions inside a local system
 by our formulation. The center of mass of a local system,
 therefore, cannot be identified from the local system
 itself, except the fact that it is at the origin of the coordinates.

Moreover, just as we have seen in the previous section, we know
 that, not only the time coordinates $t_{(H_{n\ell},\HH_{n\ell})}$
 and $t_{(H_{mk},\HH_{mk})}$, but also the space coordinates 
$x_{(H_{n\ell},\HH_{n\ell})}\in \R^3$ and 
$x_{(H_{mk},\HH_{mk})}\in\R^3$ of these two local systems are
 independent mutually. Thus the space-time coordinates 
$(t_{(H_{n\ell},\HH_{n\ell})},x_{(H_{n\ell},\HH_{n\ell})})$ and 
$(t_{(H_{mk},\HH_{mk})},x_{(H_{mk},\HH_{mk})})$ are independent
 between two different local systems $(H_{n\ell},\HH_{n\ell})$ and 
$(H_{mk},\HH_{mk})$. In particular, insofar as the systems are
 considered as quantum-mechanical ones, there is no relation between 
their centers of mass. In other words, the center of mass of any 
local system cannot be identified by other local systems
 quantum-mechanically.

Summing these two considerations, we conclude: 
\begin{namelist}{888}
\item[(1)] The center of mass of a local system $(H_{n\ell},\HH_{n\ell})$
 cannot be identified {\it quantum-mechani-cally} by any local system 
$(H_{mk},\HH_{mk})$ including the case 
$(H_{mk},\HH_{mk})=(H_{n\ell},\HH_{n\ell})$.

\item[(2)] There is no {\it quantum-mechanical} relation between any two
 local coordinates 
$(t_{(H_{n\ell},\HH_{n\ell})},$ $x_{(H_{n\ell},\HH_{n\ell})})$ and 
$(t_{(H_{mk},\HH_{mk})},x_{(H_{mk},\HH_{mk})})$ of two different
 local systems $(H_{n\ell},\HH_{n\ell})$ and $(H_{mk},\HH_{mk})$.
\end{namelist}
\MP

\F
Utilizing these properties of the centers of mass and the
 coordinates of local systems, we may make any postulates concerning
\begin{namelist}{888}
\item[(1)] the motions of the {\it centers of mass} of various local
 systems,
\end{namelist}
 and
\begin{namelist}{888}
\item[(2)] the relation between two local coordinates of any
 two local systems.
\end{namelist}
\MP

\F
In particular, we may impose {\it classical}
 postulates on them as far as the
 postulates are consistent in themselves. 
\MP

Thus we assume an arbitrary but fixed transformation:
\beq
y_2=f_{21}(y_1)\label{eqTr}
\ene
between the coordinate systems $y_j=(y_j^\mu)_{\mu=0}^3=(y_j^0,y_j^1,y_j^2,y_j^3)=(ct_j,x_j)$ for $j=1,2$,
 where $c$ is the speed of light in vacuum and $(t_j,x_j)$ is
 the space-time coordinates of the local system
 $L_j=(H_{n_j\ell_j},\HH_{n_j\ell_j})$. 
We regard these coordinates $y_j=(ct_j,x_j)$ as {\it classical}
coordinates, when we consider 
the motions of centers of mass and the relations of
 coordinates of various local systems.
We can now postulate the general principle of relativity on
 the physics of the centers of mass:

\MP

\hyphenation{centers}

{\bf Axiom 4.} 
The laws of physics which control the relative motions of the
 centers of mass of local systems are covariant under the change of
 the coordinates from
 $(ct_{(H_{mk},\HH_{mk})},$ $x_{(H_{mk},\HH_{mk})})$ to 
$(ct_{(H_{n\ell},\HH_{n\ell})},x_{(H_{n\ell},\HH_{n\ell})})$
 of the reference frame local systems for any
 pair $(H_{mk},\HH_{mk})$ and $(H_{n\ell},\HH_{n\ell})$ of
 local systems.

\MP

We note that this axiom is
 consistent with the Euclidean metric adopted for the 
quantum-mechanical coordinates inside a local system, 
because Axiom 4 is concerned with classical motions of the centers
 of mass {\it outside} local systems, and we are dealing here with
 a different aspect of nature from the quantum-mechanical one 
{\it inside} a local system.

Axiom 4 implies the invariance of the distance under the change of
 coordinates between two local systems. Thus the metric tensor
 $g_{\mu\nu}(ct,x)$ which appears here satisfies the transformation rule:
\beq
g^1_{\mu\nu}(y_1)=g^2_{\alpha\beta}(f_{21}(y_1))
\frac{\partial f^\alpha_{21}}{\partial y_1^\mu}(y_1)
\frac{\partial f_{21}^\beta}{\partial y_1^\nu}(y_1),\label{eq5}
\ene
where $y_1=(ct_1,x_1)$; $y_2=f_{21}(y_1)$ is the transformation \eq{eqTr}
 in the above from $y_1=(ct_1,x_1)$ to $y_2=(ct_2,x_2)$; and 
$g_{\mu\nu}^j(y_j)$ is the metric tensor expressed in the classical
 coordinates $y_j=(ct_j,x_j)$ for $j=1,2$.

The second postulate is the principle of equivalence, which asserts
 that the classical coordinate system 
$(ct_{(H_{n\ell},\HH_{n\ell})},x_{(H_{n\ell},\HH_{n\ell})})$ is a
 local Lorentz system of coordinates, insofar as it is concerned
 with the classical behavior of the center of mass of the local
 system $(H_{n\ell},\HH_{n\ell})$:

\MP

{\bf Axiom 5.} 
The metric or the gravitational tensor $g_{\mu\nu}$ for the center
 of mass of a local system $(H_{n\ell},\HH_{n\ell})$ in the
 coordinates 
$(ct_{(H_{n\ell},\HH_{n\ell})},x_{(H_{n\ell},\HH_{n\ell})})$ of
 itself are equal to $\eta_{\mu\nu}$, where $\eta_{\mu\nu}=0$ for 
$\mu\ne\nu$, $=1$ for $\mu=\nu=1,2,3$, and $=-1$ for $\mu=\nu=0$.

\MP

Since, at the center of mass, the classical space
 coordinates $x=0$, Axiom 5 together with the transformation rule
 \eq{eq5} in the above yields
\beq
g^1_{\mu\nu}(f^{-1}_{21}(ct_2,0))=\eta_{\alpha\beta}
\frac{\partial f^\alpha_{21}}{\partial y_1^\mu}(f^{-1}_{21}(ct_2,0))
\frac{\partial f_{21}^\beta}{\partial y_1^\nu}(f^{-1}_{21}(ct_2,0)).
\label{eq6-1}
\ene
Also by the same reason, the relativistic
 proper time 
$d\tau=\sqrt{-g_{\mu\nu}(ct,0)dy^\mu dy^\nu}$ \linebreak
$=\sqrt{-\eta_{\mu\nu}dy^\mu dy^\nu}$ at the origin of a local system
 is equal to $c$ times the quantum-mechanical proper time $dt$
 of the system.

\BP

By the fact that the classical Axioms 4 and 5 of physics
 are imposed on the centers of mass which are uncontrollable 
 quantum-mechanically, and on the relation between the coordinates of
 different, therefore quantum-mechanically non-related local systems,
 the consistency of classical relativistic Axioms 4 and 5
 with quantum-mechanical Axioms 1--3 is clear:

\BP

\begin{theorem} Axioms 1 to 5 are consistent.
\end{theorem}


\BP

\F
{\bf VI.2. Observation}

\BP

\F
Thus far, we did not mention any about the physics which
 is actually observed. We have just given two aspects of nature which
 are mutually independent. We will introduce a procedure which yields what
 we observe when we see nature. This procedure will not be
 contradictory with the two aspects of nature which we have discussed,
 as the procedure is concerned solely with ``{\it how nature looks, at
 the observer}," i.e. it is solely concerned with
 ``{\it at the place of the observer, how nature looks},"
 with some abuse of the word ``place."
 The validity of the procedure should be judged merely
 through the comparison between the observation and the prediction
 given by our procedure.

 We note that we can observe only a finite number of disjoint
 systems, say $L_1,\cdots,L_k$ with $k\ge1$ a finite integer. We
 cannot grasp an infinite number of systems at a time. Further each
 system $L_j$ must have only a finite number of elements by the same
 reason. Thus these systems $L_1,\cdots,L_k$ may be identified with
 local systems in the sense of section V.

Local systems are quantum-mechanical systems, and their coordinates
 are confined to their insides insofar as we appeal to Axioms 1--3.
 However we postulated Axioms 4 and 5 on the classical aspects of
 those coordinates, which make the local coordinates of a local
 system a classical reference frame for the centers of mass of other
 local systems. This leaves us the room to define observation as the
 {\it classical} observation of the centers of mass of local systems
 $L_1,\cdots,L_k$. We call this an observation of $L=(L_1,\cdots,L_k)$
 inquiring into sub-systems $L_1,\cdots,L_k$, where $L$ is a local
 system consisting of the particles which belong to one of the local
 systems $L_1,\cdots,L_k$.

When we observe the sub-local systems $L_1,\cdots,L_k$ of $L$,
 we observe the relations or motions among these sub-systems.
 Internally the local system $L$ behaves following the Hamiltonian
 $H_L$ associated to the local system $L$. However the actual
 observation differs from what the pure quantum-mechanical
 calculation gives for the system $L$. For example,
 when an electron is scattered by a nucleus with relative
 velocity close to that of light, the observation is different
 from the pure quantum-mechanical prediction.

%
\MP

The quantum-mechanical process inside the local system $L$ is
 described by the evolution
$$
\exp(-it_LH_L)f,
$$
where $f$ is the initial state of the system and $t_L$ is
 the local time of the system $L$. The Hamiltonian $H_L$
 is decomposed as
 follows in virtue of the local Hamiltonians $H_1,\cdots,H_k$,
 which correspond to the sub-local systems $L_1,\cdots,L_k$:
$$
H_L=H^b+T+I,\quad H^b=H_1+\cdots+H_k.
$$
Here $b=(C_1,\cdots,C_k)$ is the cluster decomposition corresponding
 to the decomposition $L=(L_1,\cdots,L_k)$ of $L$; $H^b=H_1+\cdots+H_k$
 is the sum of the internal energies $H_j$ inside $L_j$, and is an operator
 defined
 in the internal state space $\HH^b=\HH^b_1\otimes\cdots\otimes\HH^b_k$;
 $T=T_b$ denotes the intercluster free energy among the clusters
 $C_1,\cdots,C_k$ defined in the external state space $\HH_b$;
 and $I=I_b=I_b(x)=I_b(x_b,x^b)$ is the sum of the intercluster
 interactions between various two different clusters in the cluster
 decomposition $b$ (cf. the explanation after Definition 3
 in section V).

The main concern in this process would be the case that the 
clusters $C_1,\cdots,C_k$ form asymptotically bound states as
 $t_L\to\infty$, since other cases are hard to be observed along
 the process when as usual the observer's concern is upon the final state of
 the {\it bound} sub-systems $L_1,\cdots,L_k$.

The evolution $\exp(-it_LH_L)f$ then behaves asymptotically as
 $t_L\to\infty$ as follows for some bound states $g_1,\cdots,g_k$
 ($g_j\in\HH^b_j$) of local Hamiltonians $H_1,\cdots,H_k$ and for
 some $g_0$ belonging to the external state space $\HH_b$:
\beq
\exp(-it_LH_L)f\sim \exp(-it_L h_b)g_0
\otimes\exp(-it_LH_1)g_1\otimes\cdots\otimes \exp(-it_LH_k)g_k,
\quad k\ge 1,
\label{eq6}
\ene
where $h_b=T_b+I_b(x_b,0)$. It is easy to see that
 $g=g_0\otimes g_1\otimes\cdots\otimes g_k$ is given by
$$
g=g_0\otimes g_1\otimes\cdots\otimes g_k=\Omega_b^{+\ast} f
=P_b \Omega_b^{+\ast} f,
$$
provided that the decomposition of the evolution $\exp(-it_LH_L)f$
 is of the simple form as in \eq{eq6}. Here $\Omega_b^{+\ast}$ is the
 adjoint operator of a canonical wave operator (\cite{[De]})
 corresponding to the
 cluster decomposition $b$:
$$
\Omega_b^+=s{\mbox{-}}\lim_{t\to\infty}\exp(itH_L)\cdot
\exp(-ith_b)\otimes\exp(-itH_1)\otimes\cdots\otimes\exp(-itH_k)P_b,
$$
where $P_b$ is the eigenprojection onto the eigenspace of the
 Hamiltonian $H^b=H_1+\cdots+H_k$. The process \eq{eq6} just describes
 the quantum-mechanical process inside the local system $L$, and
 does not specify any meaning related with observation up to the
 present stage.

To see what we observe at actual observations, let us reflect
a process of observation of scattering phenomena. We note that the
 observation of scattering phenomena is concerned with
 their initial and final stages by what the scattering itself means.
 At the final stage of observation of scattering processes,
 the quantities observed are firstly the points hit by the scattered
 particles on the screen stood against them. If the circumstances
 are properly set up, one can further indicate the momentum
 of the scattered particles at the final stage to the extent
 that the uncertainty principle allows. Consider, e.g.,
 a scattering process of an electron by a nucleus. Given the
 magnitude of initial momentum of an electron relative to the
 nucleus, one can infer the magnitude of momentum of the electron
 at the final stage as being equal to the initial one by the law
 of conservation of energy, since the electron and the nucleus are
 far away at the initial and final stages so that the potential
 energy between them can be neglected compared to the
 relative kinetic energy. The direction of momentum at the final
 stage can also be indicated, up to the error due to the
 uncertainty principle, by setting a sequence of slits toward
 the desired direction at each point on the screen so that the
 observer can detect only the electrons scattered to that direction.
 The magnitude of momentum at initial stage can be selected in
 advance by applying a uniform magnetic field to the electrons,
 perpendicularly to their momenta, so that they circulate around
 circles with the radius proportional to the magnitude of momentum,
 and then by setting a sequence of slits midst the stream of those
 electrons. The selection of magnitude of initial momentum
 makes the direction of momentum ambiguous due to the
 uncertainty principle, since the sequence of slits lets the
 position of electrons accurate to some extent.
 To sum up, the sequences of slits at the initial and final stages
 necessarily require to take into account the uncertainty
 principle so that some ambiguity remains in the observation.
 
However, in the actual observation of a {\it single} particle,
 we {\it have to decide} at which point on the screen the 
particle hits and which momentum the particle has, using
 the prepared apparatus like the sequence of slits located at each 
point on the screen. Even if we impose an interval for the 
observed values, we {\it have to assume} that the boundaries of 
the interval are sharply designated. These are the assumption 
which we always impose on ``observations" implicitly. That is to say, 
we idealize the situation in any observation or in any measurement
 of a single particle so that the observed values for each particle
 are sharp for both of the configuration and momentum. In this 
sense, the values observed actually for each particle must be 
classical, where the {\it a priori} indefiniteness and errors
 associated with any measurement are all included. We have then necessary and sufficient conditions
 to make predictions about the differential cross section, as we 
will see in section VI.2.1.

 Summarizing, we observe just the classical quantities for each 
particle at the final stage of all observations. In other words, 
 even if we cannot know the values actually,
we have to {\it presuppose} that the values observed for each 
particle have sharp values, where all errors associated with
measurement are included.
 We can apply to this fact the remark stated in the 
third paragraph of this section about the possibility of defining 
observation as that of the {\it classical} centers of mass of 
local systems, and may assume that the actually observed values 
follow the classical Axioms 4 and 5. Those sharp values actually 
observed for each particle give, when summed over the large 
number of particles, the probabilistic nature of physical 
phenomena, i.e. that of scattering phenomena.

Theoretically, the quantum-mechanical, probabilistic
 nature of scattering processes is described by differential
 cross section, defined as the square of the absolute value
 of the scattering amplitude gotten from scattering operators
 $S_{bd}=W_b^{+\ast} W_d^- $, where $W_b^\pm$ are usual wave
 operators. Given the magnitude of the initial momentum of the
 incoming particle and the scattering angle, the differential
 cross section gives a prediction about the probability at which
 point and to which direction on the screen each particle hits
 on the average. However, as we have remarked, the idealized 
point on the screen hit by each particle and the scattering angle 
given as an idealized difference between the directions of the 
initial and final momenta of each particle have sharp values, and 
the observation at the final stage is {\it classical}. We are 
then required to correct these classical observations by 
taking into account the classical relativistic effects with those 
classical quantities, e.g., with the configuration and the momentum 
of each particle.
\BP

\F
{\bf VI.2.1.} As the first step of the relativistic modification of the
 scattering process, we consider the scattering amplitude 
$\SSS(E,\theta)$, where $E$ denotes the energy level of the scattering
 process and $\theta$ is a parameter describing the direction of the
 scattered particles. Following our remark made in the previous
 paragraph, we make the following postulate on the scattering
 amplitude observed in actual experiment:
\BP

{\bf Axiom 6.$\mbox{\bf 1}^\prime$.} When one observes the final stage of scattering
 phenomena, the total energy $E$ of the scattering process should be
 regarded as a classical quantity and is replaced by a relativistic
 quantity, which obeys the relativistic change of coordinates from
 the scattering system to the observer's system. 
\BP

Since it is not known much about $\SSS(E,\theta)$ in the many body
 case, we consider an example of the two body case. Consider a
 scattering phenomenon of an electron by a Coulomb potential 
$Ze^2/r$, where $Z$ is a real number, $r=|x|$, and $x$ is the
 position vector of the electron relative to the scatterer. We assume
 that the mass of the scatterer is large enough compared to that of
 the electron and that $|Z|/137\ll 1$. Then quantum mechanics gives
 the differential cross section in a Born approximation:
$$
\frac{d\sigma}{d\Omega}=|{\cal S}(E,\theta)|^2 
\approx \frac{Z^2e^4}{16E^2\sin^4(\theta/2)},
$$
where $\theta$ is the scattering angle and $E$ is the total
 energy of the system of the electron and the scatterer.
 We assume that the observer is stationary with respect to
 the center of mass of this system of an electron and the
 scatterer. Then, since the electron is far away from the
 scatterer after the scattering and the mass of the scatterer
 is much larger than that of the electron, we may suppose that
 the energy $E$ in the formula in the above can be replaced by the
 {\it classical} kinetic energy of the electron by Axiom 6.$1^\prime$.
 Then, assuming that the speed $v$ of the electron relative to
 the observer is small compared to the speed $c$ of light 
in vacuum and denoting the rest mass of the electron by $m$,
 we have by Axiom 6.$1^\prime$ that $E$ is observed to have the 
following relativistic value:
$$
E'=c\sqrt{p^2+m^2c^2}-mc^2
=\frac{mc^2}{\sqrt{1-(v/c)^2}}-mc^2\approx 
\frac{mv^2}{2\sqrt{1-(v/c)^2}},
$$
where $p=mv/\sqrt{1-(v/c)^2}$ is the relativistic momentum of
 the electron. Thus the differential cross section should be
 observed approximately equal to
\beq
\frac{d\sigma}{d\Omega}
\approx \frac{Z^2e^4}{4m^2v^4\sin^4(\theta/2)}(1-(v/c)^2).\label{eq7}
\ene
This coincides with the usual relativistic prediction obtained
 from the Klein-Gordon equation by a Born approximation. See 
\cite{[Ki]}, p.297, for a case which involves the spin of the electron.

Before proceeding to the inclusion of gravity in the general $k$ cluster
 case, we review this two body case. We note that the two body case
 corresponds to the case $k=2$, where $L_1$ and $L_2$ consist of
 single particle, therefore the corresponding Hamiltonians $H_1$
 and $H_2$ are zero operators on $\HH^0={\mbox{\bf C}}=$ the complex
 numbers. The scattering amplitude ${\cal S}(E,\theta)$ in this 
case is an integral kernel of the scattering matrix 
${\widehat S}={\cal F}S{\cal F}^{-1}$, where $S=W^{+\ast}W^-$ is
 a scattering operator; 
$W^\pm=s$-$\lim_{t\to\pm\infty}\exp(itH_L)\exp(-itT)$ are
 wave operators ($T$ is negative Laplacian for short-range
 potentials under an appropriate unit system, while it has
 to be modified when long-range potentials are included);
 and $\cal F$ is Fourier transformation so that 
${\cal F}T{\cal F}^{-1}$ is a multiplication operator by 
$|\xi|^2$ in the momentum representation $L^2(\R^3_\xi)$.
 By definition, $S$ commutes with $T$. This makes ${\widehat S}$
 decomposable with respect to $|\xi|^2={\cal F}T{\cal F}^{-1}$. Namely,
 for {\it a.e.} $E>0$, there is a unitary operator ${\cal S}(E)$
 on $L^2(S^2)$, $S^2$ being two dimensional sphere with radius
 one, such that for {\it a.e.} $E>0$ and $\omega\in S^2$
$$
({\widehat S}h)(\sqrt{E}\omega)
=\left({\cal S}(E)h(\sqrt{E}\cdot)\right)(\omega),
\quad h\in L^2(\R^3_\xi)
=L^2((0,\infty),L^2(S^2_\omega),|\xi|^2d|\xi|).
$$
Thus ${\widehat S}$ can be written as ${\widehat S}=\{{\cal S}(E)\}_{E>0}$.
 It is known \cite{[I-Ki]} that ${\cal S}(E)$ can be expressed as 
$$
({\cal S}(E)\varphi)(\theta)=\varphi(\theta)
-2\pi i \sqrt{E}\int_{S^2}
{\cal S}(E,\theta,\omega)\varphi(\omega)d\omega
$$
for $\varphi\in L^2(S^2)$. The integral kernel 
${\cal S}(E,\theta,\omega)$ with $\omega$ being 
the direction of initial wave, is the scattering amplitude
 ${\cal S}(E,\theta)$ stated in the above and
 $|{\cal S}(E,\theta,\omega)|^2$ is called differential
 cross section. These are the most important quantities in physics
 in the sense that they are the {\it only} quantities which can be
 observed in actual physical observation. 

The energy level $E$ in the previous example thus corresponds
 to the energy shell $T=E$, and the replacement of $E$ by $E'$
 in the above means that $T$ is replaced by a {\it classical
 relativistic} quantity $E'=c\sqrt{p^2+m^2c^2}-mc^2$. We have then
 seen that the calculation in the above gives a correct relativistic
 result, which explains the actual observation.

Axiom 6.$1^\prime$ is concerned with the observation of the final stage
 of scattering phenomena. To include the gravity into our
 consideration, we extend Axiom 6.$1^\prime$ to the intermediate process
 of quantum-mechanical evolution. The intermediate process cannot
 be an object of any {\it actual} observation, because the
 intermediate observation would change the process itself,
 consequently the result observed at the final stage would be
 altered. Our next Axiom 6.1 is an extension of Axiom 6.$1^\prime$ from
 the {\it actual} observation to the {\it ideal} observation in
 the sense that Axiom 6.1 is concerned with such invisible
 intermediate processes and modifies the {\it ideal} intermediate
 classical quantities by relativistic change of coordinates. The
 spirit of the treatment developed below is to trace the
 quantum-mechanical paths by ideal observations so that the
 quantities will be transformed into classical quantities at
 each step, but the quantum-mechanical paths will {\it not} be altered
 owing to the {\it ideality} of the observations. The classical
 Hamiltonian obtained at the last step will be ``requantized" to
 recapture the quantum-mechanical nature of the process, therefore
 the ideality of the intermediate observations will be realized
 in the final expression of the propagator of the observed system.


\BP

\F
{\bf VI.2.2.} With these remarks in mind, we return to the general
 $k$ cluster case, and consider a way to include gravity
 in our framework.

In the scattering process into $k\ge1$ clusters, what we observe
 are the centers of mass of those $k$ clusters $C_1,\cdots,C_k$,
 and of the combined system $L=(L_1,\cdots,L_k)$. In the example
 of the two body case of section VI.2.1, only the combined
 system $L=(L_1,L_2)$ appears due to $H_1=H_2=0$, therefore the
 replacement of $T$ by $E'$ is concerned with the free energy
 between two clusters $C_1$ and $C_2$ of the combined system 
$L=(L_1,L_2)$. 

Following this treatment of $T$ in the section VI.2.1, we
 replace $T=T_b$ in the exponent of 
$\exp(-it_Lh_b)=\exp(-it_L(T_b+I_b(x_b,0)))$ on the right hand
 side of the asymptotic relation \eq{eq6} by the relativistic kinetic
 energy $T'_b$ among the clusters $C_1,\cdots,C_k$ around the
 center of mass of $L=(L_1,\cdots,L_k)$, defined by
\beq
T'_b=\sum_{j=1}^k\left(c\sqrt{p_j^2+m_j^2c^2}-m_jc^2\right).
\label{eq8}
\ene
Here $m_j>0$ is the rest mass of the cluster $C_j$, which involves
 all the internal energies like the kinetic energies inside $C_j$
 and the rest masses of the particles inside $C_j$, and $p_j$ is
 the relativistic momentum of the center of mass of $C_j$ inside
 $L$ around the center of mass of $L$. For simplicity, we assume
 that the center of mass of $L$ is stationary relative to the
 observer. Then we can set in the exponent of 
$\exp(-it_L(T'_b+I_b(x_b,0)))$
\beq
t_L=t_O,\label{eq10}
\ene
where $t_O$ is the observer's time.

 For the factors $\exp(-it_LH_j)$ on the right hand side of \eq{eq6},
 the object of the {\it ideal} observation is the centers of
 mass of the $k$ number of clusters $C_1,\cdots,C_k$. These are
 the ones which now require the relativistic treatment. Since we
 identify the clusters $C_1,\cdots,C_k$ as their centers of mass
 moving in a classical fashion, $t_L$ in the exponent of
 $\exp(-it_LH_j)$ should be replaced by $c^{-1}$ times
 the classical relativistic proper time at the origin of the local
 system $L_j$, which is equal to the quantum-mechanical local time
 $t_j$ of the sub-local system $L_j$. By the same reason and by the
 fact that $H_j$ is the internal energy of the cluster $C_j$
 relative to its center of mass, it would be justified to
 replace the Hamiltonian $H_j$ in the exponent of $\exp(-it_jH_j)$
 by the classical relativistic energy {\it inside} the cluster
 $C_j$ around its center of mass
\beq
H'_j=m_jc^2, \label{eq9}
\ene
where $m_j>0$ is the same as in the above.

Summing up, we arrive at the following postulate, which has the
 same spirit as in Axiom 6.$1^\prime$ and includes Axiom 6.$1^\prime$ as a special
 case concerned with actual observation:
\BP

{\bf Axiom 6.1.} In either actual or ideal observation, the
 space-time coordinates $(ct_L,x_L)$ and the four momentum
 $p=(p^\mu)=(E_L/c,p_L)$ of the observed system $L$ should be
 replaced by classical relativistic quantities, which are transformed
 into the classical quantities $(ct_O,x_O)$ and $p=(E_O/c,p_O)$ in
 the observer's system $L_O$ according to the relativistic change
 of coordinates specified in Axioms 4 and 5. Here $t_L$ is the local
 time of the system $L$ and $x_L$ is the internal space coordinates
 inside the system $L$; and $E_L$ is the internal energy of the system
 $L$ and $p_L$ is the momentum of the center of mass of the system $L$.
\BP

In the case of the present scattering process into $k$ clusters,
 the system $L$ in this axiom is each of the local systems $L_j$
 $(j=1,2,\cdots,k)$ and $L$.

We continue to consider the $k$ centers of mass of the clusters
 $C_1,\cdots,C_k$. At the final stage of the scattering process,
 the velocities of the centers of mass of the clusters
 $C_1,\cdots,C_k$ would be steady, say $v_1,\cdots,v_k$, 
relative to the observer's system. Thus, according to Axiom 6.1, 
the local times $t_j$ $(j=1,2,\cdots,k)$ in the exponent of 
$\exp(-it_jH'_j)$, which are equal to $c^{-1}$ times the
 relativistic proper times at the origins $x_j=0$ of the
 local systems $L_j$, are expressed in the observer's time
 coordinate $t_O$ by
\beq
t_j=t_O\sqrt{1-(v_j/c)^2}\approx t_O\left(1-v_j^2/(2c^2)\right),
\quad j=1,2,\cdots,k,\label{eq11}
\ene
where we have assumed $|v_j/c|\ll 1$ and used Axioms 4 and 5 
to deduce the Lorentz transformation:
$$
t_j=\frac{t_O-(v_j/c^2)x_O}{\sqrt{1-(v_j/c)^2}},\quad 
x_j=\frac{x_O-v_jt_O}{\sqrt{1-(v_j/c)^2}}.
$$
(For simplicity, we wrote the Lorentz transformation for the
 case of 2-dimensional space-time.)

Inserting \eq{eq8}, \eq{eq10}, \eq{eq9} and \eq{eq11} into the right-hand side of
 \eq{eq6}, we obtain a classical approximation of the evolution:
\beq
\exp\left(-it_O[(T'_b+I_b(x_b,0)+H'_1+\cdots+H'_k)
-(m_1v_1^2/2+\cdots+m_kv_k^2/2)]\right)\label{eq12}
\ene
under the assumption that  $|v_j/c|\ll 1$ for all $j=1,2,\cdots,k$.

 What we want to clarify is the final stage of the scattering
 process. Thus as we have mentioned, we may assume that
 all clusters $C_1,\cdots,C_k$ are far
 away from any of the other clusters and moving almost in steady
 velocities $v_1,\cdots,v_k$ relative to the observer. We denote by
 $r_{ij}$ the distance between two centers of mass of the clusters
 $C_i$ and $C_j$ for $1\le i<j\le k$. Then, according to our spirit
 that we are observing the behavior of the centers of mass of the
 clusters $C_1,\cdots,C_k$ in {\it classical} fashion following Axioms
 4 and 5, the clusters $C_1,\cdots,C_k$ can be regarded to have
 gravitation among them. This gravitation can be calculated if we
 assume Einstein's field equation, $|v_j/c|\ll 1$, and certain
 conditions that the gravitation is weak (see \cite{[M]}, section 17.4),
 in addition to our Axioms 4 and 5. As an approximation of the 
 first order, we obtain the gravitational potential of Newtonian
 type for, e.g., the pair of the clusters 
$C_1$ and $U_1=\bigcup_{i=2}^kC_i$:
$$
-G\sum_{i=2}^km_1m_i/r_{1i},
$$
where $G$ is Newton's gravitational constant.

Considering the $k$ body classical problem for the $k$ clusters
 $C_1,\cdots,C_k$ moving in the sum of these gravitational fields,
 we see that the sum of the kinetic energies of $C_1,\cdots,C_k$
 and the gravitational potentials among them is constant by the
 classical law of conservation of energy:
$$
m_1v_1^2/2+\cdots+m_kv_k^2/2-G\sum_{1\le i<j\le k}m_im_j/r_{ij}
={\mbox{constant}}.
$$
Assuming that $v_j\to v_{j\infty}$ as time tends to infinity, we
 have constant $=m_1v_{1\infty}^2/2+\cdots+m_kv_{k\infty}^2/2$.
 Inserting this relation into \eq{eq12} in the above, we obtain the
 following as a classical approximation of the evolution \eq{eq6}:
\beq
\exp\left(-it_O\left[T'_b+I_b(x_b,0)
+\sum_{j=1}^k(m_jc^2-m_jv_{j\infty}^2/2)
-G\sum_{1\le i<j\le k}m_im_j/r_{ij}\right]\right).
\label{eq13}
\ene
What we do at this stage are {\it ideal} observations, and these
 observations should not give any sharp classical values. Thus
 we have to consider \eq{eq13} as a {\it quantum-mechanical evolution}
 and we have to recapture the quantum-mechanical feature of the
 process. To do so we replace $p_j$ in $T'_b$ in \eq{eq13} by a
 quantum-mechanical momentum $D_j$, where $D_j$ is a differential
 operator $-i\frac{\partial}{\partial x_j}
=-i\left(\frac{\partial}{\partial x_{j1}},
\frac{\partial}{\partial x_{j2}},
\frac{\partial}{\partial x_{j3}}\right)$ with respect 
to the 3-dimensional coordinates $x_j$ of the center of mass
 of the cluster $C_j$. Thus  the actual process should be
 described by \eq{eq13} with $T'_b$ replaced by a quantum-mechanical
 Hamiltonian
$$
\tT_b=\sum_{j=1}^k\left(c\sqrt{D_j^2+m_j^2c^2}-m_jc^2\right).
$$
This procedure may be called ``requantization," and is summarized
 as the following axiom concerning the ideal observation.
\BP

{\bf Axiom 6.2.} In the expression describing the classical process
 at the time of the {\it ideal} observation, the intercluster momentum
 $p_j=(p_{j1},p_{j2},p_{j3})$ should be replaced by a quantum-mechanical
 momentum $D_j=-i\left(\frac{\partial}{\partial x_{j1}},
\frac{\partial}{\partial x_{j2}},\frac{\partial}{\partial x_{j3}}\right)$.
 Then this gives the evolution describing the intermediate 
{\it quantum-mechanical} process.

\BP

We thus arrive at an approximation for a quantum-mechanical
 Hamiltonian including gravitational effect up to a constant term,
 which depends on the system $L$ and its decomposition into
 $L_1,\cdots,L_k$, but not affecting the quantum-mechanical
 evolution, therefore can be eliminated:
\SP

$$
\tH_L=\tT_b+I_b(x_b,0)-G\sum_{1\le i<j\le k}m_im_j/r_{ij}\qqq\qqq\qqq\qqq
$$
\vskip-8pt
\beq
=\sum_{j=1}^k\left(c\sqrt{D_j^2+m_j^2c^2}-m_jc^2\right)+I_b(x_b,0)
-G\sum_{1\le i<j\le k}m_im_j/r_{ij}.
\label{eq14}
\ene
\SP

\F
 We remark that the gravitational terms here come from the substitution
 of local times $t_j$ to the time $t_L$ in the factors $\exp(-it_L H_j)$
 on the right-hand side of \eq{eq6}. This form of Hamiltonian in \eq{eq14} is
 actually used in \cite{[Li]} with $I_b=0$ to explain the stability and
 instability of cold stars of large mass, showing the effectiveness
 of the Hamiltonian.

 Summarizing these arguments from \eq{eq6} to \eq{eq14}, we have obtained
 the following {\it interpretation} of the observation of the
 quantum-mechanical evolution: To get our prediction for the
 observation of local systems $L_1,\cdots,L_k$, the
 quantum-mechanical evolution of the combined local system
 $L=(L_1,\cdots,L_k)$
$$
\exp(-it_LH_L)f
$$
should be replaced by the following evolution, in the
 approximation of the first order under the assumption that
 $|v_j/c|\ll 1$ $(j=1,2,\cdots,k)$ and the gravitation is weak,
\beq
(\exp(-it_O\tH_L)\otimes 
\underbrace{I\otimes\cdots\otimes I}_{\scriptsize k \ {\mbox{factors}}})P_b
\Omega_b^{+\ast}f,
\label{eq15}
\ene
provided that the original evolution $\exp(-it_LH_L)f$ decomposes
 into $k$ number of clusters $C_1,\cdots,C_k$ as $t_L\to\infty$ in
 the sense of \eq{eq6}. Here $b$ is the cluster decomposition
 $b=(C_1,\cdots,C_k)$ that corresponds to the decomposition
 $L=(L_1,\cdots,L_k)$ of $L$; $t_O$ is the observer's time; and
\beq
\tH_L={\widetilde T}_b+I_b(x_b,0)-G\sum_{1\le i<j\le k}m_im_j/r_{ij}
 \label{eq16}
\ene
is the relativistic Hamiltonian inside $L$ given by \eq{eq14}, which
 describes the motion of the centers of mass of the clusters
 $C_1,\cdots,C_k$. 

We remark that \eq{eq15} may produce a bound state combining 
$C_1,\cdots,C_k$ as $t_O\to\infty$ therefore for all $t_O$, 
due to the gravitational potentials in the exponent. Note that
 this is not prohibited by our assumption that $\exp(-it_LH_L)f$
 has to decompose into $k$ clusters $C_1,\cdots,C_k$, because the
 assumption is concerned with the original Hamiltonian $H_L$ but
 not with the resultant Hamiltonian $\tH_L$.
\BP

\hyphenation{dif-fer-en-tial}
\hyphenation{small}
\hyphenation{since}
\hyphenation{stated}

 Extending our primitive assumption Axiom 6.$1^\prime$, which was valid for
 an example stated in section VI.2.1, we have arrived at a
 relativistic Hamiltonian $\tH_L$, which would describe approximately
 the intermediate process, under the assumption
 that the gravitation is weak and the velocities of the particles
 are small compared to $c$, by using the Lorentz transformation.
 We note that, since we started our argument from the asymptotic
 relation \eq{eq6}, which is concerned with the final stage of
 scattering processes, we could assume that the velocities
 of particles are almost steady relative to the observer
 in the correspondent
 classical expressions of the processes, therefore we could appeal to
 the Lorentz transformations when performing the change of coordinates
 in the relevant arguments. 

 The final values of scattering amplitude should
 be calculated by using the Hamiltonian $\tH_L$. Then they would
 explain actual observations. This is our prediction for the observation
 of relativistic quantum-mechanical phenomena including the effects by
 gravity and quantum-mechanical forces.

 In the example discussed in section VI.2.1, this approach gives
 the same result as \eq{eq7} in the approximation of the first order,
 showing the consistency of our spirit (see \cite{Ki2}).


\hyphenation{physics}

\BP

\newpage

\vskip 24pt

\large
\noindent
{\bf VII. Existence of Local Motion}

\vskip12pt

\normalsize

\F
We are in a position to see how the stationary nature of the universe and the existence of local motion and hence local time are compatibly incorporated into our formulation.

\BP

\noindent
{\bf VII.1. G\"odel's theorem}
\BP

\F
Our starting point is the incompleteness theorem proved by
 G\"odel \cite{G}. It states that any consistent
 formal theory that can
 describe number theory includes an infinite number of undecidable
 propositions. The physical world includes at least
 natural numbers, and it is described by a system of words, which
 can be translated into a formal physics theory. The theory of
 physics, if consistent, 
 therefore includes an undecidable proposition, i.e. a proposition
 whose correctness cannot be known by human beings until one finds
 a phenomenon or observation that supports the proposition or
 denies
 the proposition. Such propositions exist infinitely according to
 G\"odel's theorem. Thus human beings, or any other finite entity,
 will never be able to reach a ``final" theory that can express
 the totality of the phenomena in the universe.

Thus we have to assume that any human observer sees a part
 or subsystem $L$ of the universe and never gets the total
 Hamiltonian $H$ in \eq{eq1} by his observation. Here the
 total Hamiltonian $H$ is an {\it ideal} Hamiltonian
that might be gotten by ``God." In other words, a consequence
 from G\"odel's theorem is that the Hamiltonian that an
 observer assumes with his observable universe
 is a part $H_L$ of $H$. Stating explicitly, 
the consequence from G\"odel's theorem is the
 following proposition
\beq
H=H_L+I+H_E,\q H_E\ne 0,\label{G2}
\ene
where $H_E$ is an unknown Hamiltonian describing
 the system $E$ exterior to the realm of the observer,
 whose existence, i.e. $H_E\ne 0$, is assured by G\"odel's
 theorem. This unknown system $E$ includes
all that is unknown to the observer. 
E.g., it might contain particles which
 exist near us but have not been discovered yet,
 or are unobservable for some reason at the time of
 observation.
The term $I$ is an unknown interaction between
 the observed system $L$ and the unknown system $E$. 
Since the exterior system $E$ is assured to exist
by G\"odel's theorem, the interaction $I$ does not vanish:
 In fact assume $I$ vanishes. Then the observed system $L$
 and the exterior system $E$ do not interact, which is
 the same as that the exterior system $E$ does not exist
 for the observer. 
On the other hand,
assigning the so-called G\"odel number to each
 proposition in number theory, G\"odel constructs 
undecidable
propositions in number theory by a diagonal argument, 
which shows that any consistent formal
 theory has a region exterior to the knowable world
(see \cite{G}).
Thus the observer 
 must be able to construct a proposition by G\"odel's
 procedure that proves
 $E$ exists, which means $I\ne 0$.
By the same reason, $I$ is not
 a constant operator:
\beq
I \ne \mbox{constant operator}.\label{G3}
\ene
For suppose it is a constant operator. Then
 the systems $L$ and $E$ do not change no matter how far or
 how near they are located because the interaction
 between $L$ and $E$ is a constant operator.
 This is the same situation as that the interaction does not
 exist, thus reduces to the case $I=0$ above.

We now arrive at the following observation:
For an observer, the observable universe is a part $L$ 
of the total universe and it looks as though it follows the
 Hamiltonian $H_L$, not following the total Hamiltonian $H$.
 And the state of the system $L$ is described by a part
 $\phi(\cdot,y)$ of the state $\phi$ of the total universe,
 where $y$ is an unknown coordinate of system $L$ inside
 the total universe, and $\cdot$ is the variable controllable
 by the observer, which we will denote by $x$.
\BP


\F
{\bf VII.2. Local Time Exists}
\BP

\F
In the following argument, we assume an exact relation:
\beq
H\phi=0 \label{eq1-prime}
\ene
instead of \eq{eq1}, for simplicity.

Assume now, as is usually expected under condition \eq{eq1-prime},
 that there is no
 local time of $L$, i.e. that the state $\phi(x,y)$
 is an eigenstate of the local Hamiltonian $H_L$ for
 some $y=y_0$ and a real number $\mu$:
\beq
H_L\phi(x,y_0)=\mu\phi(x,y_0).\label{G4}
\ene
Then from \eq{G2}, \eq{eq1-prime} and \eq{G4} follows that
\beq
&&0=H\phi(x,y_0)
=H_L\phi(x,y_0)+I(x,y_0)\phi(x,y_0)+H_E\phi(x,y_0)\nonumber\\
&&\ \hskip5pt=(\mu+I(x,y_0))\phi(x,y_0)+H_E\phi(x,y_0).\label{G5}
\ene
Here $x$ varies over the possible positions of the particles
 inside
 $L$. On the other hand, since $H_E$ is the Hamiltonian
 describing the system $E$ exterior to $L$, it does not
 affect the variable $x$ and acts only on the variable $y$.
 Thus $H_E\phi(x,y_0)$ varies as a bare function $\phi(x,y_0)$
 insofar as the variable $x$ is concerned.
Equation \eq{G5} is now written: For all $x$
\beq
H_E\phi(x,y_0)=-(\mu+I(x,y_0))\phi(x,y_0).\label{G6}
\ene
As we have seen in \eq{G3}, the interaction $I$ 
is not a constant operator and varies when $x$
 varies\footnote{Note that G\"odel's theorem
 applies to any fixed $y=y_0$ in \eq{G3}. Namely,
 for any position $y_0$ of the system $L$ in the
 universe, the observer must be able to know
 that the exterior system $E$ exists because
 G\"odel's theorem is a universal statement
 valid throughout the universe.
 Hence $I(x,y_0)$ is not a constant operator
 with respect to
 $x$ for any fixed $y_0$.},
 whereas the action
 of $H_E$ on $\phi$ does not.
 Thus there is a nonempty set of points $x_0$
 where $H_E\phi(x_0,y_0)$ and $-(\mu+I(x_0,y_0))\phi(x_0,y_0)$
 are different, and \eq{G6} does not hold at such points
 $x_0$. If $I$ is assumed to be continuous in the variables
 $x$ and $y$, these points $x_0$ constitutes a set of
 positive measure. This then implies that our assumption
 \eq{G4} is wrong. Thus a subsystem $L$ of the universe cannot
 be a bound state with respect to the observer's Hamiltonian
 $H_L$. This means that the system $L$ is observed as
 a non-stationary system, therefore there must be observed
 a motion inside the system $L$. This proves that the
 ``time" of the local system $L$ {\it exists for the
 observer} as a measure of motion, whereas the total
 universe is stationary and does not have ``time."

\newpage

\BP

\F
{\bf VII.3. A refined argument}
\BP

\F
To show the argument in section VII.2 more explicitly,
we consider a simple case of
$$
H=\frac{1}{2}\sum_{k=1}^N
h^{ab}(X_k)p_{ka} p_{kb}+V(X).
$$
Here $N$ $(1\le N\le \infty)$ is the number of particles
 in the universe, $h^{ab}$ is a three-metric, 
$X_k\in R^3$ is the position of the $k$-th particle, 
$p_{ka}$ is a functional derivative corresponding to
 momenta of the $k$-th particle, and
$V(X)$ is a potential. The configuration
 $X=(X_1,X_2,\cdots,X_N)$ of total particles is decomposed
 as $X=(x,y)$ accordingly to if the $k$-th particle
 is inside $L$ or not, i.e. if the $k$-th particle is
 in $L$, $X_k$ is a component of $x$ and if not it is
 that of $y$. $H$ is decomposed as follows:
$$
H=H_L+I+H_E.
$$
Here $H_L$ is the Hamiltonian of a subsystem $L$ that
 acts only on $x$, $H_E$ is the Hamiltonian describing the
 exterior $E$ of $L$ that acts only on $y$, and
$I=I(x,y)$ is the interaction between the systems $L$ and $E$.
 Note that $H_L$ and $H_E$ commute.

\BP

\begin{theorem} Let $P$ denote the eigenprojection
onto the space of all bound states of $H$.
Let $P_L$ be the eigenprojection for $H_L$. Then we have
\begin{equation}
(1-P_L)P \ne 0,\label{G7}
\end{equation}
unless the interaction $I=I(x,y)$ is a constant with
 respect to $x$ for any $y$.
\end{theorem}
\MP

\noindent
{\it Proof}. 
Assume that \eq{G7} is incorrect. Then we have
$$
P_LP=P.
$$
Taking the adjoint operators on the both sides, we then have
$$
PP_L=P.
$$
Thus $[P_L,P] = P_LP - PP_L = 0$.
 But in generic this does not hold because
$$
[H_L,H] = [H_L, H_L+I+H_E] = [H_L,I]\ne 0,
$$
unless $I(x,y)$ is equal to a constant with respect to $x$.
 Q.E.D.

\BP

\noindent
{\bf Remark.} In the context of section V,
 the theorem implies the following: 
$$
(1-P_L)P \UU \ne \{ 0 \},
$$
where $\UU$ is a Hilbert space consisting of all
 possible states $\phi$ of the total universe.
 This relation implies that there is a vector $\phi\ne 0$
 in $\UU$ which satisfies $H\phi=\lambda \phi$ for
 a real number $\lambda$ while $H_L \Phi \ne \mu \Phi$
 for any real number $\mu$, where $\Phi=\phi(\cdot,y)$
 is a state vector of the subsystem $L$ with an appropriate
 choice of the position $y$ of the subsystem. Thus the space
generated by $\phi(\cdot,y)$'s when $y$ varies is non-trivial in the
sense of Definition 1 in section V, which proves
for the universe $\phi$ 
 that any local system $L$ is non-trivial, and hence
proves the existence of local time for any local system
of the universe $\phi$. Thus we have at least one 
stationary universe $\phi$ where
every local system has its local time.
\BP

\BP

\end{document}